\documentclass[
 reprint,
nofootinbib,
 amsmath,amssymb,
 aps,
floatfix 
]{revtex4-2}

\usepackage{graphicx}
\usepackage{dcolumn}
\usepackage{bm}
\usepackage[colorinlistoftodos]{todonotes}
\usepackage[export]{adjustbox}%
\usepackage{natbib}
\usepackage{lineno}
\usepackage{amsthm}

\usepackage{xcolor}
\colorlet{rcolor}{black}  

\DeclareMathOperator*{\argmax}{arg\,max}

\begin{document}

\title{Rationalizing risk aversion in science}

    \author{Kevin Gross}
    \email{krgross@ncsu.edu}
    \affiliation{Department of Statistics \\ North Carolina State University \\ Raleigh, NC USA}
    
    \author{Carl T. Bergstrom}
    \email{cbergst@u.washington.edu}
    \affiliation{Department of Biology \\ University of Washington \\ Seattle, WA USA \\} 

	\date{\today}

\begin{abstract}
Scientific research requires taking risks, as the most cautious approaches are unlikely to lead to the most rapid progress. Yet much funded scientific research plays it safe and funding agencies bemoan the difficulty of attracting high-risk, high-return research projects. Why don't the incentives for scientific discovery adequately impel researchers toward such projects? Here we adapt an economic contracting model to explore how the unobservability of risk and effort discourages risky research. The model considers a hidden-action problem, in which the scientific community must reward discoveries in a way that encourages effort and risk-taking while simultaneously protecting researchers' livelihoods against the vicissitudes of scientific chance.  Its challenge when doing so is that incentives to motivate effort clash with incentives to motivate risk-taking, because a failed project may be evidence of a risky undertaking but could also be the result of simple sloth.  As a result, the incentives needed to encourage effort actively discourage risk-taking. Scientists respond by working on safe projects that generate evidence of effort but that don't move science forward as rapidly as riskier projects would. A social planner who prizes scientific productivity above researchers' well-being could remedy the problem by rewarding major discoveries richly enough to induce high-risk research, but scientists would be worse off for it. Because the scientific community is approximately self-governing and constructs its own reward schedule, the incentives that researchers are willing to impose on themselves are inadequate to motivate the scientific risks that would best expedite scientific progress. 

\end{abstract}

\maketitle

\section*{Introduction}

Scientific inquiry is a risky business.  Every experiment, every analysis, every collaboration entails embarking on a path whose destination is uncertain and whose terminus could be a dead end \cite{franzoni2023uncertainty}. Not all projects are equally risky however. In choosing what to work on, scientists have the latitude to embrace risk or to shy away from it. High risk can bring high return; some of the biggest scientific advances arise from risky projects, and credit accrues to investigators accordingly \cite{rzhetsky2015choosing, stephan2015economics}.  Excessive caution hampers scientific progress, and so scientists, funders of science, and the public all have an interest in encouraging researchers to pursue risky projects.  Yet, formal and anecdotal evidence alike suggest that investigators shy away from taking the big risks that may generate the most productive science \cite{foster2015tradition, franzoni2022funding}. Even efforts to solicit risky research proposals have met with middling success \cite{franzoni2022funding, wagner2013evaluating}. These observations beg the question: why don't the incentive structures in science impel researchers to pursue sufficiently high-risk research?

In this paper, we use a mathematical model to study how scientists' willingness to pursue risky projects is affected by the non-observability of their actions. By ``the non-observability of their actions'', we mean that scientists are not directly rewarded for their effort or for the degree of risk they assume in their research.  
Instead, investigators are rewarded for the scientific advances that their research yields.  However, the uncertain nature of science means that a scientist's productivity is only stochastically related to the effort they expend and the scientific risk they take on.  Thus, in deciding how hard to work and how much scientific risk to embrace, an investigator must weigh the potential for discoveries and acclaim against the chance that their project will fail, leaving them with little to show for their efforts. 
Here, we aim to understand how the unobservable, or hidden, nature of effort and risk shape investigators' research strategies and the incentive structures within which they work.\footnote{Observability is not the only factor that affects investigators' risk preferences. Franzoni et al.\ \cite{franzoni2022funding} have recently reviewed a range of explanations for why science funders and scientists who seek funding may be averse to scientific risk. Elsewhere, we have considered how the {\em ex ante} nature of grant peer review can deter investigators from proposing projects that their colleagues would consider risky \cite{gross2021ex}.} 

The claim that effort and risk-taking are unobservable to those who ultimately determine how an investigator's work is rewarded---namely, to the scientific community---is the central premise of our model.  While the unobservability of risk-taking and effort is a foundational tenet of the economics of science \cite{dasgupta1994new,stephan1996economics}, it nevertheless benefits from further justification in the present context.  Of course, scientists' actions, especially their effort, are observable to those around them---their bosses, their colleagues, their family, and their friends.  But these are not the people who ultimately determine whether scientists get the university tenure-track job or are honored by a professional society. Instead, career success accrues based on how the scientific community evaluates scientists' contributions.  Thus it is the scientific community that serves as the ultimate arbiter of investigators' professional success, and this community does not have the means (by virtue of its physical dispersion) or perhaps even the interest to monitor investigators' effort.\footnote{At first glance, it may seem that scientists' supervisors have ample say in the professional rewards those scientists receive and would be able to monitor effort if they were so inclined.  Yet in academic science supervisors primarily reward researchers based on the esteem in which those researchers' contributions are held by scientific community.  Thus, supervisors act in response to the community's judgment, not separately from it.}   Indeed, to a first approximation, scientists' careers rise and fall based on their CVs, and CVs list only scholarly outputs, not hours worked or risky undertakings gone bust.  This set of circumstances, and thus our model, pertain most directly to investigators who work individually or in small teams on projects that may take a few months or a few years, or at most a single career.  They do not describe, and thus our model does not capture, large infrastructure projects such as space exploration or high-energy physics where the need to coordinate huge teams makes effort plainly visible to the scientific community in which those teams operate.

Formally, we study investigators' risk-taking behavior using the economic framework of a hidden-action problem.  Hidden-action, or moral hazard, problems are most often used  to study the tensions that arise when a ``principal'' writes a contract to hire an ``agent'' to work on principal's behalf \cite{holmstrom1979moral,grossman1983analysis}.  In our set-up, individual investigators are the agents while the scientific community collectively plays the role of the principal.\footnote{Indeed, academic science is a self-organized pursuit, in which scientists operate within institutions---hiring norms, systems for funding, tenure standards, etc.---that they have designed for themselves and that endure only with their continued blessing.  While this is undoubtedly a simplification, it hews close enough to reality to expose fundamental tensions at work in motivating scientists to do risky research.  When necessary to avoid ambiguity, we refer to scientists acting collectively as the ``scientific community'' and individual scientists acting in their own best interest as ``scientists'', ``investigators'', or ``researchers''.}  The ``contract'' that the scientific community establishes is not an actual contract, of course, but instead represents the traditions that the community establishes for rewarding investigators for their discoveries.  In other words, we study how much prestige the scientific community assigns to a publication in one journal as opposed to another, or to a failed research attempt that generates no publication at all.

In establishing these traditions, the scientific community must collectively solve the following problem.  The community receives resources---jobs, funding, prestige, etc.--- from the public to support the scientific endeavor in exchange for the knowledge that the community produces \cite{yin2022public}.  The community seeks to distribute these resources among its members in so as to maximize researchers' individual well-being, but in doing so it is constrained in the following ways.  First, because only scientists' discoveries (e.g., publications, etc.) are observable, the community's tradition for distributing rewards can only be based on the scientific advances that researchers produce.  Second, the community is not free to bestow unlimited rewards on its members.  Instead, the total volume of rewards is linked to the public's aggregate support of science.  Third, while individual researchers are not disposed to embrace or to shy away from scientific risk for its own sake, researchers are averse to putting their livelihoods at risk in the usual way that economic actors are assumed to be averse to livelihood risk.

To sum, we model how a scientific community establishes a tradition for rewarding discoveries that motivates investigators to work hard and to take scientific risks while also protecting researchers' livelihoods from the vicissitudes of scientific chance.  
For the sake of comparison, we also solve the same contracting problem from the perspective of a (fictitious) benevolent social planner who cares most about optimizing scientific progress and who only cares secondarily about investigators' well-being. The rest of this article proceeds as follows.  We first present the model in its most general form and illustrate the main result with a numerical example.  The general model does not have enough structure to permit a complete analysis, so we describe the intuition behind the main result by analyzing a simplified version of the model.  Mathematical details and proofs appear in the appendix.  

The most closely related modeling work to this analysis seems to be Manso's study of the tension between risk-taking and effort in the context of executive compensation \cite{manso2011motivating}.   The scientific community's problem also resembles an optimal taxation problem, in the sense that the community redistributes the fruits of scientific progress in a way that maximizes researchers' individual well-being while preserving the incentive for investigators to make important discoveries \cite{laffont2002theory}.  Franzoni and colleagues \cite{franzoni2022funding} provide a recent qualitative analysis of the reasons why investigators often eschew risky research.

\section*{Mathematical model}

Our model envisions a scientific workforce that consists of a unit mass of investigators.  To get off the ground, we make two substantial assumptions.  First, we assume that all scientists are alike.  Second, we consider a one-shot setting in which investigators attempt a single study instead of building a portfolio of complementary studies. Clearly, both of these assumptions are unrealistic.  Their purpose here is to allow us to focus on the key strategic dilemma that the community faces when attempting to motivate both risk-taking and hard work. Relaxing either or both of these assumptions complicates the community's task in interesting ways that would be ripe for subsequent analysis.

When an investigator pursues a study they must decide both how risky of a project to pursue and how much effort to invest.  Let $r$ represent a study's scientific risk and let $e$ represent an investigator's effort.  We will let both $r$ and $e$ take values in $[0, 1]$, with larger values of $r$ and $e$ corresponding to greater risk and more effort, respectively.  Together, the pair $(r,e)$ forms an investigator's action.  We assume that there is no direct cost to scientific risk, in the sense that risky projects are no more or less onerous than safer ones.\footnote{In reality, it may be that riskier projects are also more onerous, which if true would exacerbate the tension studied here.}  By contrast, effort is costly to investigators, such that an investigator who expends effort $e$ incurs a disutility cost $c(e)$, with $c(0)=0$, $c'>0$, and $c''>0$.\footnote{The convexity of $c$ can be motivated by assuming that researchers obtain decreasing marginal returns to leisure, a common assumption in labor economics \cite{deaton1980economics}.} 

Each study generates an outcome that has a scientific value of $v \geq 0$.  Unpublishable outcomes contribute nothing to science and have value $0$.  Publishable outcomes have strictly positive value, with larger values of $v$ corresponding to more valuable outcomes.  Let $F(v; r, e)$ give the cumulative probability distribution of $v$ when an investigator expends effort $e$ on a study with risk $r$.  We assume that the values of published outcomes are continuously distributed with probability density $f(v;r,e) > 0$ on $v>0$, with $\int_{v>0} \, f(v; r, e) \, dv = 1 - F(0; r, e)$, where $F(0; r, e)$ is the probability of getting an unpublishable outcome.\footnote{The analysis also holds if $v$ takes a discrete distribution for publishable outcomes.}   We assume that $f(v; r, e)$ is twice differentiable in $r$ and $e$ and differentiable in $v$. 
Use $\int \, g(v) \, F(dv; r, e)$ to denote the expectation of a function $g(v)$ when the distribution of $v$ is given by $F(v; r, e)$, and let $s(r, e) = \int \, v \, F(dv; r, e)$ denote the expected scientific productivity of an investigator who takes action $(r,e)$. Because we have assumed that all investigators are alike, $s(r,e)$ also gives the aggregate per capita productivity of the scientific community when every investigator takes action $(r,e)$.  We assume that the scientific community is large enough that the stochasticity in the community's aggregate productivity is negligible.  Let $(\hat{r}, \hat{e}) = \argmax s(r,e)$ be the action that maximizes scientific productivity.  We assume $(\hat{r}, \hat{e})$ is unique.

\begin{figure}[h!]
	\begin{center}
		\includegraphics[width=\linewidth]{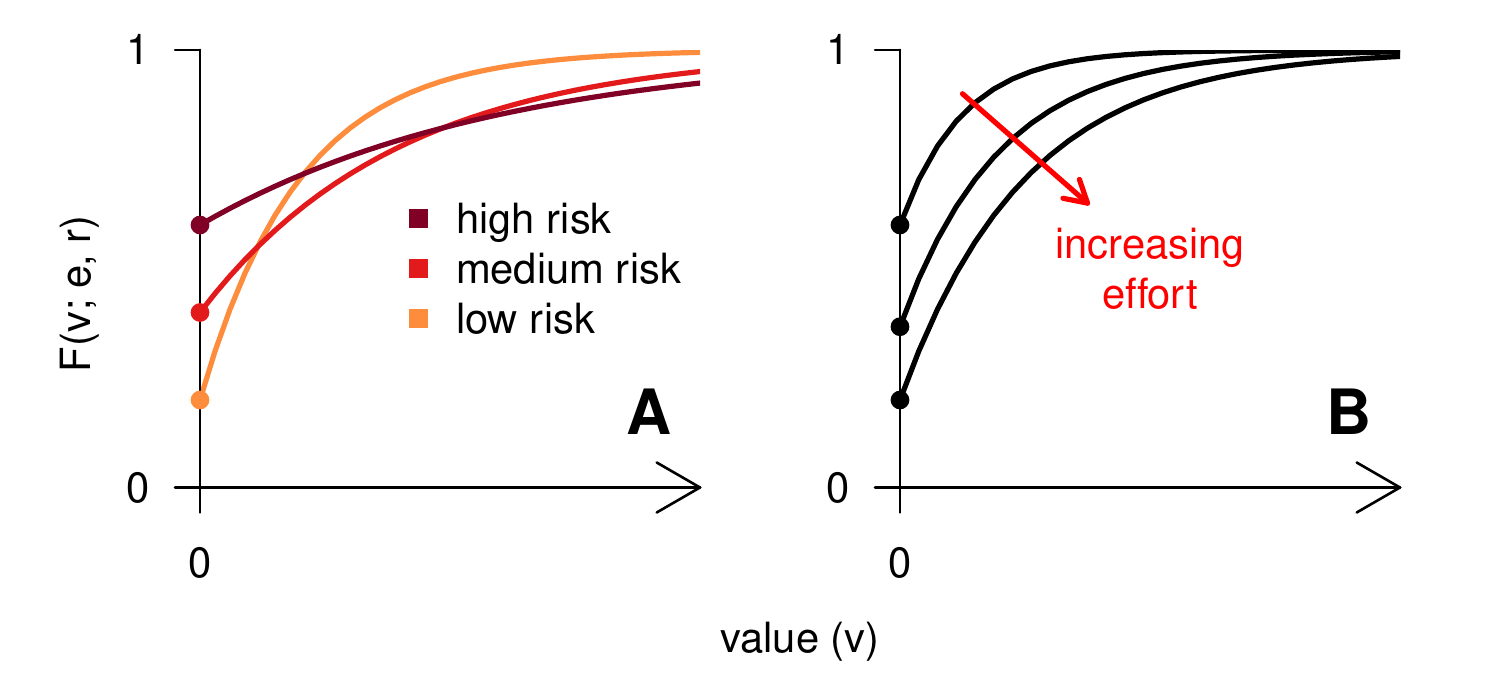}
	\end{center}
	\caption{\textbf{Risk and effort have differing effects on the distribution of a project's value.} Each panel shows the cumulative distribution function $F(v; r, e)$ of a project's value $v$.  Values at $v=0$ correspond to the probability of generating an unpublishable outcome.  A: Increasing scientific risk increases the probability of getting an unpublishable outcome but also increases the expected value of a publication conditional on publication.  B: Increasing effort decreases the probability of getting an unpublishable outcome and increases the probability of publishing an outcome with a large value.}
	\label{fig:v-cdf}
\end{figure}

Suppose that the scientific value of a study depends on effort and risk in the following way.  Consider risk first.  While there are many facets of scientific risk \cite{mandler2017benefits,veugelers2022funding, franzoni2023uncertainty}, here we assume that scientific risk has the following properties.  First an increase in risk increases the probability that an investigator will unluckily generate an unpublishable outcome and will have nothing to show for their efforts (Fig.~\ref{fig:v-cdf}A).  Because scientific risk has no natural units, we might as well equate a project's scientific risk with the probability that a study generates an unpublishable outcome when an investigator gives full effort; that is, $F(0; e=1, r) = r$.  
Second, an increase in risk increases the conditional expectation of an outcome's scientific value given that it is publishable.  Third, we assume that $s(r,e)$ is strictly concave in $r$ for any $e$.

Now consider effort. Zero effort ($e=0$) guarantees zero science ($v=0$), that is, $F(0; 0, r) = 1$.  Second, for any scientific risk and any scientific value $v$, increasing effort strictly increases the the chance of obtaining an outcome at least as valuable as $v$ (Fig.~\ref{fig:v-cdf}B).  In other words, increasing effort increases $v$ in the sense of first-order stochastic dominance.\footnote{That is, $F(v; e_1, r) > F(v; e_2, r)$ for any $r \in (0,1)$, any $v$, and any $e_1<e_2$.}  This implies that the expected scientific productivity $s(r,e)$ increases in effort.  

More technically, we also require that $F(v; r, e)$ satisfies a regularity condition called the monotone likelihood ratio property (MLRP) with respect to both $r$ and $e$.  The MLRP is standard for hidden-action problems \cite{milgrom1981good}.  The MLRP with respect to risk-taking states that among publishable outcomes a higher-valued publication provides more evidence of greater risk than a lower-valued publication provides. With respect to effort, the MLRP requires that among all studies (publishable or not) a larger value of $v$ provides at least as much evidence of greater effort than a smaller value of $v$ provides.  These conditions are discussed in more detail in the appendix.

We now turn to the problem that the scientific community faces in establishing its tradition for rewarding investigators for their discoveries.  Society provides resources to the scientific community in return for scientists' collective progress.  These resources include all forms of compensation, such as income, jobs, prestige, etc.   We assume that the total resources are directly linked to the aggregate scientific productivity \cite{yin2022public} and write the per capita resources (the community's resource ``budget'') as $B(s)$, with $B(0) = 0$ and $B' > 0$.  The resource budget represents a longstanding arrangement between the public and the scientific community that emerges over many rounds of the one-shot process that we consider.  

The scientific community distributes these resources among its members based on the discoveries that researchers generate.  Let $w(v) \geq 0$ denote the resource reward (or ``wages'') that a scientist receives in return for a scientific outcome of value $v$.  We call $w(\cdot)$ the community's ``contract'', and we assume that $w(\cdot)$ is differentiable for $v>0$. A scientist's utility increases by $u(w) \geq 0$ if they receive wage $w$.  We assume that $u(0) = 0$, $u' > 0$, and $u''<0$, so that scientists are strictly risk averse with respect to their wages. Write $\bar{u}(v) = u(w(v))$, and call $\bar{u}(v)$ the wage-utility for $v$. A scientist who chooses $(r,e)$ while facing contract $w(\cdot)$ obtains the payoff
\begin{equation}
	\pi(r, e, w(\cdot)) = \int \, \bar{u}(v) \, F(dv; r, e) - c(e).
\end{equation}
We equate $\pi(r,e,w(\cdot))$ with a researcher's well-being.  A contract $w(\cdot)$ is said to implement the action $(r,e)$ if an investigator facing $w(\cdot)$ maximizes their well-being by choosing that action.  

The scientific community's problem is to find a contract $w(\cdot)$ that maximizes its members' well-being while implementing an action that generates enough aggregate scientific productivity to justify the public's investment.  In other words, the community's problem is 
\begin{equation}
	\max_{r, e, w(\cdot)} \pi(r, e, w(\cdot))
	\label{eq:cp}
\end{equation} 
subject to the following constraints.  First, we assume that the contract must reward high-value science at least as handsomely as it rewards lower-value science.  That is, the contract must satisfy the monotonicity constraint 
\begin{linenomath}
	\begin{align}
		w(0) & \leq \lim_{v \downarrow 0} w(v) \nonumber \\
		w'(v) & \geq 0 \mbox{ for } v>0.
		\label{eq:mc}
		\tag{MC}
	\end{align} 
\end{linenomath}
Second, because the total rewards cannot exceed the resource budget, the contract and the action it implements must satisfy the budget constraint
\begin{equation}
	\int \, w(v) \, F(dv; r, e) \leq B(s(r, e)).
	\label{eq:bc}
	\tag{BC}
\end{equation}
Third, the contract must implement the action.  That is, the contract must satisfy the incentive constraint
\begin{equation}
	(r, e) \in \argmax_{r', e'} \pi(r', e', w(\cdot)).
	\label{eq:ic}
	\tag{IC}
\end{equation}
Both the effort and risk components of the IC optimize over a continuum of possible actions and thus entail a continuum of constraints.  We follow the usual procedure of analyzing instead the relaxed problem in which both components of the IC are replaced by the corresponding first-order conditions
\begin{equation}
	\frac{\partial}{\partial r} \pi(r, e,w(\cdot)) = 0
	\label{eq:rc}    
	\tag{RC}
\end{equation}
and
\begin{equation}
	\frac{\partial}{\partial e} \pi(r, e,w(\cdot)) = 0.
	\label{eq:ec}    
	\tag{EC}
\end{equation}
We call these the risk constraint (RC) and the effort constraint (EC), respectively.  While cautions apply to using a first-order condition \cite{rogerson1985first, mirrlees1999theory, chade2020no}, 
we assume that the second-order conditions needed to justify the use of the RC and EC hold for any $w(\cdot)$ that satisfies the MC. Write the action implemented by the community's optimal contract as $(\tilde{r}, \tilde{e})$.\footnote{\label{footnote:pc} Hidden-action problems also typically involve a participation constraint.  Suppose that an investigator may leave science to pursue an outside option that provides the reservation payoff $\pi_R \geq 0$.  The participation constraint (PC) would then read $\pi(r, e, w(\cdot)) \geq \pi_R$, assuming that investigators participate at indifference. To ensure that there are at least some contracts that make it worthwhile for investigators to participate in science, we assume that there is at least one action for which an investigator's payoff under the ``soft-money'' contract $w(v) = v B(s) / s$ satisfies the PC.  Under this assumption, the PC has no bite in the scientific community's problem.}

We consider two benchmark settings for comparison.  The first is the hypothetical full-information scenario in which the investigators' actions are observable.  In this scenario the community can mandate that investigators must take a particular action to receive the rewards offered by the contract.  
The full-information solution solves (\ref{eq:cp}) subject to only the MC and BC.  The second benchmark is the problem that would be faced by a hypothetical social planner whose primary objective is to maximize scientific productivity.  This planner's problem is
\begin{equation}
	\max_{r, e, w(\cdot)} s(r, e)
	\label{eq:sp}
\end{equation}
subject to the MC, BC, RC, and EC.\footnote{The PC (see footnote \ref{footnote:pc}) may have bite in the social planner's problem.  When it does, the social planner finds the contract that maximizes scientific productivity while just providing the investigators with a payoff of $\pi_R$.  Because the social planner's problem is only a benchmark and not the primary focus of this article, we do not explore the consequences of the PC for the social planner any further.}  We will see that the social planner may have several contracts available that implement the productivity-maximizing action $(\hat{r},\hat{e})$.  In these cases, we assume that the social planner is benevolent and seeks the contract that maximizes the investigators' well-being while implementing $(\hat{r},\hat{e})$. 

\section*{Analysis}

We analyze two models that differ in the effort component of the researchers' action.  In the first model, effort affects both the probability of generating a publishable outcome and the conditional distribution of a publishable outcome's scientific value.  While this is the most realistic model, it is too flexible to permit a complete analysis.  Instead we present a numerical example that illustrates the main result.  To explore the mechanism behind the main result, we then analyze a simplified model in which effort only affects the probability of generating a publishable outcome.  This simplified model has enough structure to prove the main result and allows us to explore the intuition behind it.  All numerical calculations were computed using R \cite{R}.  Full mathematical arguments appear in the appendix.

\subsection*{Model 1: Effort affects the both the probability and value of publication}

In the most general formulation, both risk and effort affect the probability of generating a publishable outcome and the scientific value of publishable outcomes.  In this set-up, risk-taking and effort are only distinguished by the fact that greater risk decreases the probability of a publishable outcome and increases the value of a publishable outcome, while greater effort increases both.   This set-up has too little structure to support a provable result about whether the community's tradition distorts scientific risk-taking away from its progress-maximizing level.  Instead of a general result, we consider a numerical example.  In this numerical example, the probability of generating a publishable outcome is given by $\sqrt{e}(1-r)$ and outcomes that are publishable have an exponentially distributed scientific value with mean $r\sqrt{e}$.  Thus scientific productivity $s(r,e) = e r (1 - r)$, and the productivity-maximizing action is $(\hat{r} = 1/2, \hat{e} = 1$). An investigator's disutility cost of effort is given by $c(e) = 0.1 e^2$ and their utility derived from wages is given by $u(w) = \sqrt{w}$.  Finally, the community's resource budget is linear in the aggregate scientific productivity, $B(s) = s$.

\begin{figure}[h!]
	\begin{center}
		\includegraphics[width=\linewidth]{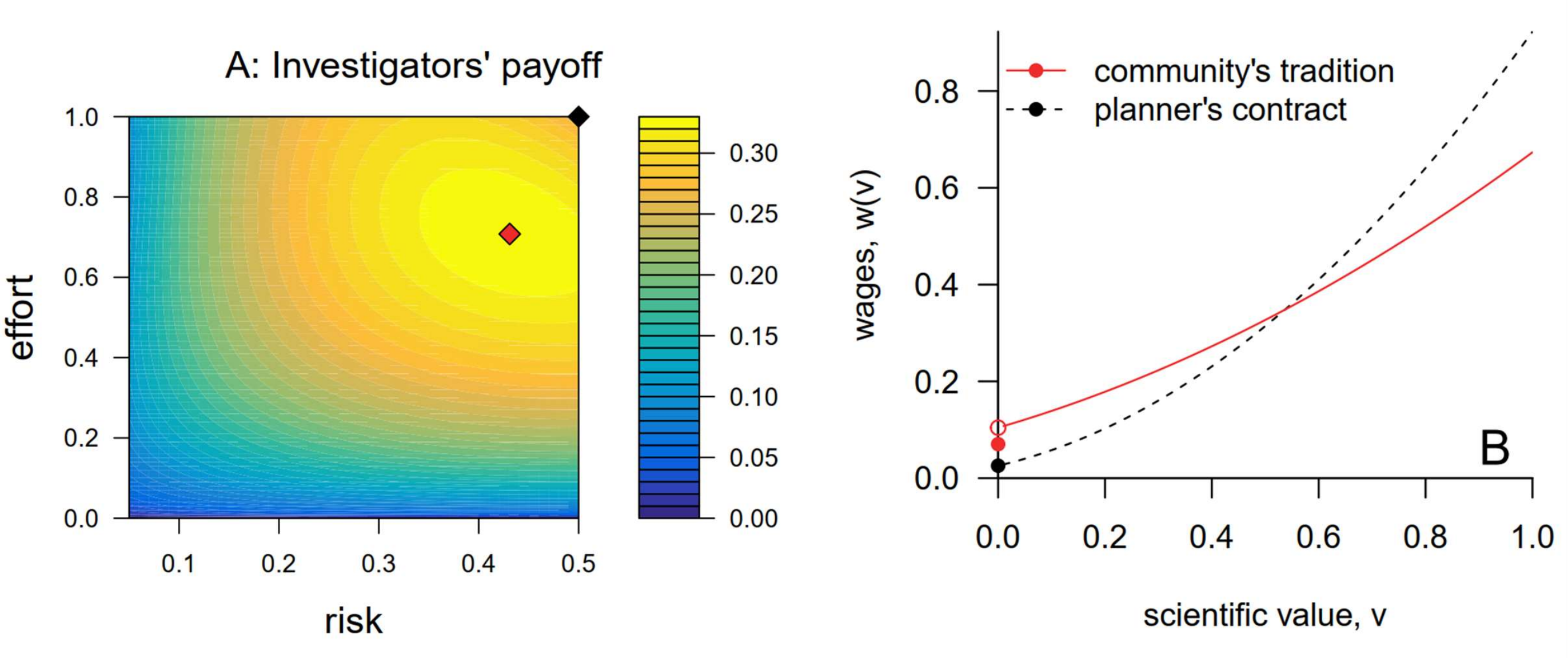}
	\end{center}
	\caption{\textbf{The scientific community's tradition decreases both risk-taking and effort away from their productivity-maximizing actions while rewarding scientific contributions more evenly.}  A: The investigators' payoff at the optimal tradition, or contract, for implementing each possible action.  The red diamond shows the action that yields the largest possible payoff and hence the community's preferred tradition.  The black diamond shows the payoff that the investigators would receive from a benevolent social planner who seeks first to maximize scientific producitivity and secondly to optimize investigators' welfare.  B: The reward tradition established by the community (red curve) and the contract that a benevolent social planner would favor (black). The community's preferred tradition distributes rewards more evenly among investigators than the the social planner's contract does. }
	\label{fig:example}
\end{figure}

In this example, the community's reward tradition distorts risk-taking and effort downwards from their productivity-maximizing values.  In Fig.~\ref{fig:example}A, this is visualized by rewriting the community's problem (eq.~\ref{eq:cp}) as 
\begin{equation}
	\max_{r,e} \max_{w(\cdot)} \pi(r, e, w(\cdot)).
	\label{eq:cp2}
\end{equation}  
Fig.~\ref{fig:example}A shows the solution to the inner maximization of eq.~\ref{eq:cp2} for each action; that is, it shows the investigator's payoff at the payoff-maximizing contract that implements each particular action.  The outer maximization in eq.~\ref{eq:cp2} is then found by locating the action that maximizes this payoff, as shown by the red diamond in Fig.~\ref{fig:example}A. This action and the corresponding reward tradition (Fig.~\ref{fig:example}B) solve the community's problem.

In this example, a benevolent social planner can institute a contract that impels researchers to take the productivity-maximizing action.  The social planner does this by solving the inner maximization of eq.~\ref{eq:cp2} for $(\hat{r}, \hat{e})$. The community's reward tradition distributes rewards more evenly among its members than the social planner's contract (Fig.~\ref{fig:example}B).  The community's tradition reduces scientific productivity relative to the social planner's contract ($s(\tilde{r}, \tilde{e}) \approx 0.174$ for the community's tradition vs.\ $s(\hat{r}, \hat{e}) =0.25$ for the social planner's contract), but it leaves the investigators better off. 

In the full-information case (when investigators' actions are observable), the community would establish a tradition that pays everyone who takes the contracted action the same wage regardless of the scientific output that their research yields.  The intuition here is clear: If actions are observable, then the community bears the scientific risk and insures investigators fully, because an investigator who takes the mandated action but produces a weak or unpublishable result has manifestly just been unlucky. In the numerical example, the community's optimal contract mandates that each investigator take the productivity-maximizing action and pays each investigator who does so a wage equal to the per capita scientific productivity ($w = 0.25$).  This gives the investigators a greater payoff ($\pi = 0.4$) than they would receive under any hidden-action scenario.

\subsection*{Model 2: Effort scales the probability of publication}

The numerical example above suggests that the scientific community's tradition distorts risk-taking and effort downwards from their productivity-maximizing values.  The downward distortion of costly effort echoes the standard result of moral-hazard models, in which the there is a tension between motivating costly effort and insuring the agents against livelihood risk \cite{holmstrom1979moral, grossman1983analysis, laffont2002theory}.  The downward distortion of risk, though, is more surprising, because risk incurs no cost \textit{per se} to the investigators.  Of course, without further investigation, it is hard to say whether this result is general or idiosyncratic to the settings of the numerical example.  To unpack the downward distortion of risk-taking, we turn to a simplified model (Model 2) in which effort affects only the probability of publication but does not affect the distribution of $v$ for publishable outcomes. More specifically, we assume that the probability of generating a publishable outcome scales linearly with $e$ and thus equals $e(1-r)$.  Let $F(v;r)$ give the cumulative denisty function (cdf) of $v$ under full effort. Then the cdf of $v$ under any effort in the simplified model is then given by $F(v;r,e) = 1 - e + eF(v;r)$.

In this simplified model, the community's tradition always distorts risk-taking downward from its productivity-maximizing level, that is, $\tilde{r} < \hat{r}$ as long as $F(v;r)$ meets a few additional mild technical conditions.  A mathematical proof appears in the appendix.  Here, we sketch the intuition behind the result.  To do so, we appeal to a standard result from moral-hazard theory, which is that the optimal contract for implementing a particular action rewards outcomes in proportion to the evidence or ``good news'' that those outcomes provide for that action \cite{holmstrom1979moral, grossman1983analysis}. 

First, note that the budget constraint BC binds (this is true even in the more general Model 1).  The intuition here is straightforward: the community doesn't leave resources on the table.  Because the BC binds, each researcher's expected reward equals the per capita (expected) scientific productivity.

Now consider an intermediate scenario in which only risk-taking is hidden but effort is observable.  In this case, the community can give the same reward to everyone who exerts the optimal effort, thus insuring the investigators against the stochasticity inherent in science.  Because investigators are fully insured, they are free to pursue the productivity-maximizing (and thus wage-maximizing) scientific risk without exposing their livelihoods to risk.  Consequently, scientific risk-taking is not distorted when effort is observable.\footnote{The situation here may seem tenuous: at full insurance, an investigator's only reason to favor the productivity-maximizing risk is that doing so maximizes her expected contribution to the aggregate productivity.  However, if the number of investigators is large, then any single investigator's contribution to the aggregate productivity will be negligible and their motivation to choose the productivity-maximizing risk will be weak. However, the community can create a strong incentive to pursue the productivity-maximizing risk by blending a small fraction ($\epsilon$)  of the social planner's wage scheme with $1-\epsilon$ of full insurance, thus only decreasing the wage payment by a small increment.}

Now suppose both risk-taking and effort are hidden.  Of course, when effort is hidden, researchers cannot be fully insured against the risk of a non-publishable outcome; if they were, they would have no incentive to work hard.  In the simplified Model 2, the only evidence of a  researcher's effort is provided by whether or not they publish.  Thus the equilibrium tradition must reward researchers who publish more handsomely than researchers who don't. This reward premium for publication forces researchers to bear some livelihood risk, decreasing their well-being.  

Yet---and here lies the crux of the matter---while a reward premium for publication is necessary to motivate effort, it simultaneously discourages scientific risk-taking by penalizing researchers who generate unpublishable results. With full insurance rendered impossible, the community must counteract the risk-discouraging effects of effort provision by rewarding higher-value publications more handsomely, thus restoring an incentive to pursue risky projects.  Yet the uneven reward tradition needed to motivate risk-taking creates its own drag on researchers' well-being by exposing them to additional livelihood risk.  Because of this drag, the community settles for a more egalitarian tradition that balances partial protection from livelihood risk against the reduced productivity of the more conservative science that this tradition implements.  Compare this with the social planner who, because they are less concerned with the researchers' well-being, can mandate a contract that is sufficiently uneven to fully counteract the risk-discouraging effects of effort provision.


To understand why the additional structure of Model 2 guarantees that risk-taking will be distorted downwards, consider the interaction between incentives for risk-taking and for effort.  Incentives for risk-taking and effort will always be in tension at $v=0$, because an unpublishable result provides ``good news'' about an investigator's risk-taking but ``bad news'' about their effort.  In Model 2, this tension at $v=0$ is the entire interaction between motivating effort and motivating risk-taking.  Thus, the need to motivate effort makes the mechanisms necessary to motive risk more costly (in the sense of decreasing researchers’ well-being) and vice versa.  In Model 1, though, it is conceivable that ``good news'' about effort and risk-taking might be strongly enough aligned for publishable outcomes ($v>0$) to counteract the tension at $v=0$, thereby making it less costly to motivate effort when investigators pursue riskier-than-optimal science.  Although this scenario seems far-fetched to us in practice, nothing in the mathematical structure of Model 1 seems to rule out the possibility.\footnote{While the simplifications in Model 2 suffice to guarantee that risk-taking is distorted downwards from its maximizing level, they likely aren't necessary.  We speculate that the downward distortion of risk-taking can be guaranteed under milder simplifications than Model 2 imposes, but articulating those conditions is also beyond the scope of this article.}

\section*{Discussion}

On the face of it, scientists as a group seem to face dilemma that they are unable to solve.  On the one hand, risky research generates the groundbreaking advances that expand our knowledge most rapidly.  On the other hand, scientists seem either unable or unwilling to devise institutions that motivate investigators to embrace the scientific risks that would lead to the most rapid progress \cite{franzoni2022funding}. 
Our analysis here suggests that this state of affairs can be explained at least in part by the interaction between two key structural elements in science: the unobservability of risk and effort on the one hand, and the self-organized nature of science on the other.  If either of these elements were reversed---if either risk and effort could be verifiably documented, or if science was governed by a hypothetical social planner with the authority to allocate prestige unilaterally---scientists could either motivate themselves or be motivated by the social planner to implement the productivity-maximizing scientific risk.  The mechanisms would be quite different, however; in the former case, scientists would be rewarded directly for their risk and effort choices, while in the latter case the social planner would heap prestige on investigators who generate the most ground-breaking discoveries while reducing the prestige awarded for more incremental advances. 

But neither option is available.  We have argued earlier about the hidden nature of risk and effort, ruling out the possibility of awarding prestige based on either.  We have said less about why a social planner does not emerge to organize (academic) science.  For one, a social planner cannot emerge because science is a specialized pursuit, and assessing the scientific value of discoveries requires a close knowledge of the field.  Thus a hypothetical social planner would be hamstrung by being forced to rely on the guidance of the investigators to determine how to value outcomes, and as such would be no more than a conduit for the investigators' collective judgements \cite{carmichael1988incentives}.  In other words, they wouldn't be a planner at all.

Why doesn't the scientific community simply adopt the scheme that a social planner would advocate?  They don't because the social planner's scheme leaves the scientists worse off, despite optimizing scientific progress.  The social planner's scheme demoralizes the investigators because it places the investigators at too great a risk of having little to show for their efforts if their scientific risk does not pay off.  The scientific community can do better by weakening, though not eliminating, the disparity in prestige awarded for groundbreaking vs.\ incremental outcomes, preserving the incentive for investigators to take on some scientific risk while also protecting their livelihoods if the scientific risk doesn't pan out.


The tension we study here helps to explain some features of the scientific ecosystem which might otherwise seem perverse.  Consider scientific funding.  Many funding bodies have programs dedicated to funding risky research.  Yet those funders also ask investigators to report the research outputs generated by prior funding awards.  While few would question that a researcher's past productivity provides useful information to reviewers evaluating new proposals, it is still worth nothing that funders discourage risk-taking when they make prior productivity a {\em de facto} requirement for subsequent funding. On its face, this practice seems to send a mixed message to investigators.  Yet such a mixed message may be a reasonable compromise for a community that can only verify a researcher's outputs.  In other words, while the scientific community may be perpetually frustrated by its inability to impel investigators to take bigger risks, this frustration is not necessarily evidence of poor institutional design; it may instead be an unavoidable consequence of information asymmetries inherent in the scientific endeavor.

Our analysis makes a number of simplifications, each of which provide an opportunity for further research.  Perhaps most substantively, we have assumed that all investigators are alike.\footnote{The assumption of worker homogeneity in common in hidden-action models of the sort that we explore here.  Indeed, as we have already noted, our model is similar to models of redistributive income taxation in which workers face uncertainty about the income that their action will yield.  Most such models in the economic literature assume workers are all alike; see Vairo \cite{vairo2024} for a literature summary and a recent approach to accommodating heterogeneity among workers.}  In science, researchers differ in many ways that affect how they design their research programs, including their abilities and their appetite for livelihood risk.  These differences are only privately known to the individual investigators, at least initially.  In contracting theory, private differences among agents further complicate the principal's task through the phenomenon of ``adverse selection'', in which the principal's inability to know agents' type results in additional distortions away from efficient outcomes.  Moreover, the effects of adverse selection compound when investigators also make hidden-action decisions \cite{laffont2002theory}.  Without further study, we can only speculate about how these interactions may play out in science, although it seems likely that adverse selection complicates the task of designing a reward scheme that motivates researchers to pursue the type of research at which they excel.  

Second, our assumption that researchers are alike makes it easy to determine the scientific community's objective, because the contract that is best for one researcher is also best for all.  In reality, different types of researchers will prefer different reward schemes.  Understanding how the community resolves these differences requires an additional understanding of the internal politics of science that eludes the authors.

Our model also simplifies the community's task by considering only a one-off setting.  In reality, of course, researchers build their careers and their reputation through a series of projects, thus allowing investigators to build a research portfolio that may include both risky endeavors and safer bets.  Further, the act of doing science is not as simple as our model envisions.  In reality, projects evolve, and investigators routinely make operational decisions throughout a project's lifetime that steer it towards its conclusion.  Perhaps one of the most helpful skills in science is the ability to make the proverbial lemonade from lemons, that is, to generate useful science from a risky effort gone bust.  Finally, in the long-run, a researcher's output begins to yield information about her abilities relative to those of other investigators, introducing an interaction between one's career stage and the adverse selection mentioned above that only complicates matters further \cite{holmstrom1999managerial}. All these realities intrude on the simple setting we have analyzed here in ways that will require future work to unravel. 

These caveats notwithstanding, the dynamic that we have explored here is inescapable.  In deciding how to reward discoveries, the scientific community must contend with the fact that reward schemes that motivate effort inherently discourage scientific risk-taking, and vice versa. Because the community must motivate both effort and scientific risk-taking, and because effort is costly to investigators, the community inevitably establishes a tradition that encourages more conservative science than would be optimal for maximizing scientific progress, even when risky research is no more onerous than safer lines of inquiry.

\section*{Acknowledgments}
We thank Lones Smith, Kyle Myers, and Zack Brown for their helpful comments on an early draft of this paper. This work was supported by NSF awards SMA 19-52343 and SMA 19-52069 to KG and CTB respectively, and by Templeton World Charity Foundation grant 20650 to CTB. KG thanks the Department of Biology at the University of Washington for visitor support.

\bibliography{risk}

\newpage 

\renewcommand\theequation{A.\arabic{equation}}  
\setcounter{equation}{0}   

\renewcommand{\thesubsection}{A.\arabic{subsection}}
\renewcommand{\thesubsubsection}{A.\arabic{subsection}.\arabic{subsubsection}}

\onecolumngrid
\section*{Appendix}

This appendix contains a proof of our main result, details about the monotone likelihood ratio property, and numerical examples.  We begin by analyzing the full-information solution to the community's problem, and then proceed to analyze Models 1 and 2 in turn.  Details for the numerical example follow.

Throughout, we take the standard step \cite{laffont2002theory} of re-writing the scientific community's problem in terms of the wage-utility schedule $\bar{u}(v) = u(w(v))$. Let $h(\cdot) = u^{-1}$ give the wages needed to achieve utility $u$, with $h(0) = 0$, $h' > 0$, and $h'' > 0$.  To simplify notation, we write the scientists' payoff $\pi$ as a function of their action $(r, e)$ and the wage utility schedule $\bar{u}(v)$. We also use $f(v; r, e)$ to denote the density of $F(v; r, e)$ with respect to a measure that sums counting measure at $v=0$ and Lebesgue measure on $\mathcal{R}$. Use subscripts to denote partial derivatives, e.g., $f_r(v; r, e) = \partial f(v; r, e) / \partial r$.  Unless otherwise noted, all integrals are with respect to $v \geq 0$.  Write the community's (per capita) resource budget under action $(r, e)$ as $\bar{B}(r,e) = B(s(r,e))$.

We take the usual step of separating the problem into two sub-problems \cite{grossman1983analysis}, as shown in eq.~\ref{eq:cp2}.  The first sub-problem, or inner maximization in eq.~\ref{eq:cp2}, solves for the contract that maximizes the community's payoff when implementing any action $(r,e)$, subject to the same constraints as the full problem.   Write the contract that solves this sub-problem as $\bar{u}^{(r,e)}$, that is, 
\begin{equation}
 \bar{u}^{(r,e)}(\cdot) = \argmax_{\bar{u}(\cdot)} \pi(r, e, \bar{u}(\cdot)).
 \label{eq:cp-sub}
\end{equation}
This sub-problem has a linear objective and a convex constraint space, so any solution that satisfies the Kuhn-Tucker conditions will be a global optimum \cite[Thm.\ M.K.3]{mascolell1995microeconomic}.   The second sub-problem, or outer maximization in eq.~\ref{eq:cp2}, finds the action that maximizes the payoff $\pi(r, e, \bar{u}^{(r,e)}(\cdot))$, that is, it solves 
\begin{equation}
 \max_{r,e} \pi(r, e, \bar{u}^{(r,e)}(\cdot)).
\end{equation}

\subsection{Monotone likelihood ratio property}

The monotone likelihood ratio property (MLRP) is a regularity condition on $F(v; r, e)$.  For risk, the MLRP states that a higher-valued publication provides more evidence of greater risk than a lower-valued publication provides.  In notation, this writes as 
\begin{equation}
    \label{eq:mlrp-r}
    \dfrac{\partial}{\partial v} \left[ \dfrac{f_r(v; r, e)}{f(v; r, e)}\right] > 0
\end{equation}
for $v>0$.  An equivalent statement of the MLRP for $r$ is that for any two values $v_2 > v_1 > 0$, any two risk levels $r_2 > r_1$, and any $e$, then 
\begin{displaymath}
    \frac{f(v_1; r_2, e)}{f(v_1; r_1, e)} < \frac{f(v_2; r_2, e)}{ f(v_2; r_1; e)};
\end{displaymath}
this expression accords more closely with the name ``monotone likelihood ratio''.  Note that the MLRP for risk does not extend to $v=0$.  Indeed, because $F_r(0;r,e) + \int_{v>0} f_r(v;r,e) \, dv = 0$, $F_r(0; r, e)> 0$ and eq.~\ref{eq:mlrp-r} together imply $\lim_{v \downarrow 0} f_r(v; r, e) < 0$.  Thus 
\begin{equation}
    \label{eq:mlrp-0}
  \lim_{v \downarrow 0}\dfrac{f_r(v; r, e)}{f(v; r, e)} < 0 < \dfrac{F_r(0; r, e)}{F(0; r, e)}.
\end{equation}

With respect to effort, the MLRP says that a larger value of $v$ provides at least as much evidence of greater effort than a smaller value of $v$ does.  In notation, this condition writes as 
\begin{equation}
    \dfrac{\partial}{\partial v} \left[ \dfrac{f_e(v; r, e)}{f(v; r, e)}\right] \geq 0.
    \label{eq:mlrp-e}
\end{equation} 
Crucially, the MLRP for effort extends to $v=0$.  The fact that the MLRP for effort extends to $v=0$ while the MLRP for risk does not creates an unavoidable tension between the incentives for risk-taking and the incentives for effort that stymies the community's ability to motivate both efficiently.

\subsection{Full information}

Under full information, the community's problem is only constrained by the monotonicity constraint MC and budget constraint BC. To show this formally, assume that the MC will hold.  Using $\beta \geq 0$ as the multiplier on the BC, the Lagrangian for sub-problem \ref{eq:cp-sub} writes as 
\begin{linenomath}
\begin{displaymath}
 L(\bar{u}^{(r,e)}(\cdot), \beta; r, e) = \int \, \bar{u}^{(r,e)} \, F(dv; r, e) - c(e) + \beta \left[ \bar{B}(r, e) -  \int \, h(\bar{u}^{(r,e)})  \, F(dv; r, e) \right]. 
\end{displaymath}
\end{linenomath}
with the slackness condition $\beta \left[ \bar{B}(r, e) -  \int \, h(\bar{u}^{(r,e)}(v))  \, F(dv; r, e) \right]  = 0$.  Set $\partial L / \partial \bar{u}^{(r,e)}(v) = 0$ and differentiate pointwise to give
\begin{linenomath}
\begin{equation*}
0 = f(v; r, e) \left[ 1 - \beta h'(\bar{u}^{(r,e)}(v)) \right].
\end{equation*}
\end{linenomath}
Thus $h''>0$ implies $\bar{u}^{(r,e)}(v) = \bar{u}^{(r,e)}$ is constant for all $v$, and $h'>0$ implies $\beta = 1 / h'( \bar{u}^{(r,e)}) > 0$, thus the BC binds.  Note that $\beta$ (the shadow price of the budget constraint) equals $u'(w)$, the marginal utility of wages at equilibrium; a similar interpretation of $\beta$ will hold when the action is hidden. If $\bar{u}^{(r,e)}(v)$ is constant and the BC binds, then every investigator receives an equal wage $w =  \bar{B}(r, e)$.  Thus the community contracts on the action that maximizes $u(\bar{B}(r,e)) - c(e)$.

In general, it is not necessarily the case that the community will contract on the progress-maximizing action. Indeed, the particular action that the community contracts upon will depend on the particular forms of $u(\cdot)$, $B(\cdot)$, $s(r, e)$, and $c(\cdot)$.  However, because risk-taking is not intrinsically onerous, the community will contract on the progress-maximizing level of scientific risk for any particular effort level.  That is, for any $e$, $\argmax_r u(\bar{B}(r,e)) = \argmax_r s(r,e)$. This follows from $u'>0$ and $B'>0$.

\subsection{Analysis of Model 1}

In Model 1, there are several binding constraints, so we must check that the constraint qualification holds.  It is straightforward to show that the only way in which the qualification would not hold is if $f_e(v;r,e) / f_r(v;r,e)$ is constant for all $v$.\footnote{The constraint qualification holds if the constraints are linearly independent.  Assuming the BC will bind, the qualification is violated if we can find a contract $\bar{u}(v)$ and values $\eta_1$, $\eta_2$, and $\eta_3$, not all 0, such that $\eta_1 h'(\bar{u}(v))f(v; r, e) + \eta_2 f_e(v; r, e) + \eta_3 f_r(v;r,e) = 0$ for all $v \geq 0$.  Integrating over $v\geq0$ gives $\eta_1 \int \, h'(\bar{u}(v))F(dv; r, e) + \eta_2 \int \, F_e(dv; r, e) +  \eta_3 \int \, F_r(dv; r, e)$ = 0, from which $\int \, F_e(dv; r, e) = \int \, F_r(dv; r, e) = 0$ implies $\eta_1 = 0$.  Thus the constraint qualification is only violated if $f_e(v;r,e) / f_r(v;r,e)$ is constant for all $v$.}.  We assume that this will not be the case; it is easy to verify as much for the numerical example.

Use $\lambda$ and $\mu$ as the multipliers for EC, and RC, respectively, and assume that the MC will hold. The Lagrangian for sub-problem \ref{eq:cp-sub} writes as 
\begin{linenomath}
\begin{align}
 L(\bar{u}^{(r,e)}(\cdot), \beta, \lambda, \mu; r, e) & = \int \, \bar{u}^{(r,e)}(v) \, F(dv; r, e) - c(e) + \beta \left[ \bar{B}(r, e) -  \int \, h(\bar{u}^{(r,e)}(v))  \, F(dv; r, e) \right]  \nonumber \\ & + \lambda\left[ \int \, \bar{u}^{(r,e)}(v) \, F_e(dv; r, e) - c'(e) \right] + \mu \int \, \bar{u}^{(r,e)}(v) \, F_r(dv; r, e)
 \label{eq:full-lagrangian}
\end{align}
\end{linenomath}
with $\beta>0$ and the slackness condition $\beta \left[ \bar{B}(r, e) -  \int \, h(\bar{u}^{(r,e)}(v))  \, F(dv; r, e) \right]  = 0$.  We have reversed the sign of $\lambda$ and $\mu$ from the way they are typically written to ease the forthcoming interpretation.  As before, set $\partial L / \partial \bar{u}^{(r,e)}(v) = 0$ to give 
\begin{linenomath}
\begin{equation}
0 = f(v; r, e) - \beta h'(\bar{u}^{(r,e)}(v))  f(v; r, e)  + \lambda f_e(v; r, e) + \mu f_r(v; r, e).
\label{eq:dLdu}
\end{equation}
\end{linenomath}
Integrate with respect to $v$ obtain
\begin{linenomath}
\begin{equation*}
0 = \int \, F(dv; r, e)  - \beta \int \, h'(\bar{u}^{(r,e)}(v)) F(dv; r, e) + \lambda \int \, F_e(dv; r, e) + \mu \int \, F_r(dv; r, e).
\end{equation*}
\end{linenomath}
Now use $\int \, F(dv; r, e) = 1$ and $\int \, F_e(dv; r, e) = \int \, F_r(dv; r, e) = 0$ to obtain
\begin{linenomath}
\begin{equation*}
\beta^{-1} = \int \, h'(\bar{u}^{(r,e)}(v)) F(dv; r, e)
\end{equation*}
\end{linenomath}
from which $\beta>0$ and the BC binds, as before.

Now divide eq.~\ref{eq:dLdu} by $f(v; r, e)$ and rearrange to yield
\begin{linenomath}
\begin{equation}
\beta h'(\bar{u}^{(r,e)}(v)) - 1 = \lambda \dfrac{f_e(v; r, e)}{f(v; r, e)} + \mu \dfrac{f_r(v; r, e)}{f(v; r, e)}.
\label{eq:stareqn}
\end{equation}
\end{linenomath}
By $\beta>0$, the MC, and $h''>0$, the LHS of eq.~\ref{eq:stareqn} must be non-decreasing in $v$. We can rule out $\lambda = \mu = 0$, because this would give a constant wage for all $v$, and a constant wage cannot implement costly effort.  By our assumptions about the MLRP, $f_e(v; r, e)/f(v; r, e)$ is non-decreasing in $v$ for all $v$, while $f_r(v; r, e)/f(v; r, e)$ is not monotone (it is decreasing in $v$ at $v=0$, and strictly increasing in $v$ for $v>0$).  Thus we must have $\lambda > 0$,\footnote{Suppose not.  If $\lambda = 0$, there is no choice of $\mu \neq 0$ that would make the RHS of eq.~\ref{eq:stareqn} non-decreasing in $v$, because $f_r(v; r, e)/f(v; r, e)$ is not monotone.  If $\lambda < 0$, we would need $\mu<0$ to make the RHS of eq.~\ref{eq:stareqn} non-decreasing at $v=0$, yet $\lambda <0$ and $\mu<0$ would make the RHS of eq.~\ref{eq:stareqn} strictly decreasing in $v=0$ for $v>0$.  Note also that $\lambda$ is the shadow price of the EC; thus $\lambda>0$ indicates that (reasonably) the payoff decreases as the EC tightens.} and the EC binds.  

Without further assumptions on the structure of $F(v; r, e)$, we cannot take the analysis any further.  Most notably, the sign of $\mu$ is ambiguous, and we will see that establishing $\mu>0$ is key to the proof below that the community distorts risk downwards at equilibrium in Model 2.  

\subsection{Analysis of Model 2}

Here, we show that in Model 2 the community's tradition is guaranteed to distort risk-taking downwards from its productivity-maximizing level.  The basic path of the proof is to show that when effort does not affect the conditional distribution of $v$ for publishable outcomes, $f_e(v; r, e)/f(v; r, e)$ is constant for $v>0$, which in turn implies $\mu>0$ by eq.~\ref{eq:stareqn}.  By applying the Envelope Theorem it can be shown that $\mu>0$ implies $s_r(\tilde{r}, \tilde{e})>0$.  Because we have assumed that $s(r,e)$ is strictly concave in $r$ for any $e$, it then follows that $\tilde{r} < \argmax_r s(\tilde{e}, r)$. Finally, because $s(r,e)$ is separable in $e$ and $r$ in Model 2, it follows that $\argmax_r s(r, e) = \hat{r}$ for any $e$, and thus $\tilde{r} < \hat{r}$. 

Let $f(v;r) = f(v; e = 1, r)$ give the density of $v$ under full effort ($e=1$).  In Model 2, the full density of $v$ can be written 
\begin{linenomath}
\begin{equation*}
    f(v; r, e) = \begin{cases} 1 - e(1 - f(0; r)) & v = 0 \\
                     e f(v; r) & v > 0 \end{cases}
\end{equation*}    
\end{linenomath}
Thus, the likelihood ratio with respect to $e$ is constant for $v>0$, that is, $f_e(v; r, e)/f(v; r, e) = 1/e$.  By eq.~\ref{eq:stareqn}, it then follows that $\mu>0$ and the RC binds.\footnote{We have already argued that the LHS of eq.~\ref{eq:stareqn} must be non-decreasing in $v$ for all $v \geq 0$.  Further, $\bar{u}^{(r,e)}(v)$ must be strictly increasing in $v$ for at least some $v>0$; otherwise, the contract could only implement the minimal scientific risk $r=0$. Thus the LHS of eq.~\ref{eq:stareqn} must be strictly increasing in $v$ for some $v>0$ as well.  But if $f_e(v; r, e)/f(v; r, e)$ is constant for $v>0$, this requires $\mu>0$.}

Write the investigators' payoff at the optimal contract for implementing action $(r,e)$ as $\Pi(r,e) = \pi(r, e, \bar{u}^{(r,e)}(\cdot))$.  Apply the Envelope Theorem \cite[p.\ 456]{simon1994mathematics} to give
\begin{linenomath}
\begin{align}
 \dfrac{\partial \Pi(r,e)}{\partial r} & = \int \, \bar{u}^{(r,e)}(v) \, F_r(dv; r, e) + \beta \left[ \bar{B}_r(r, e) -  \int \, h(\bar{u}^{(r,e)}(v))  \, F_r(dv; r, e) \right] \nonumber  \\ & + \lambda \left[ \int \, \bar{u}^{(r,e)}(v) \, F_{re}(dv; r, e) \right] + \mu \int \, \bar{u}^{(r,e)}(v) \, F_{rr}(dv; r, e). 
\label{eq:dLdr}
\end{align}
\end{linenomath}
The first term on the RHS in \ref{eq:dLdr} vanishes because $\bar{u}^{(r,e)}(v)$ implements $r$, and hence $\frac{\partial}{\partial r}\int \, \bar{u}^{(r,e)}(v) \, F(dv; r, e) = 0$ by RC.  The key element of Model 2 is that term in square brackets that multiplies $\lambda$ in eq.~\ref{eq:dLdr} also vanishes:
\begin{linenomath}
\begin{align}
  \int \, \bar{u}^{(r,e)}(v) \, F_{re}(dv; r, e) & = \dfrac{\partial^2}{\partial e \, \partial r} \int \, \bar{u}^{(r,e)}(v) \, F(dv; r, e) \nonumber \\
  & = \dfrac{\partial^2}{\partial e \, \partial r} \left[ \bar{u}^{(r,e)}(0)(1 - e) + e \int \, \bar{u}^{(r,e)}(v) \, F(dv; r)\right]\nonumber \\
  & =  \dfrac{\partial}{\partial r} \int \, \bar{u}^{(r,e)}(v) \, F(dv; r) \nonumber \\
  & = 0
  \label{eq:cross-partial}
\end{align}
\end{linenomath}
where the last equality follows from RC.\footnote{To see this, RC implies $0 = \frac{\partial}{\partial r}\int \, \bar{u}^{(r,e)}(v) \, F(dv; r, e)= \frac{\partial}{\partial r}\left[ \bar{u}^{(r,e)}(0)(1 - e) + e \int \, \bar{u}^{(r,e)}(v) \, F(dv; r)\right] = e\frac{\partial}{\partial r} \int \, \bar{u}^{(r,e)}(v) \, F(dv; r)$, and thus $\frac{\partial}{\partial r} \int \, \bar{u}^{(r,e)}(v) \, F(dv; r) = 0$ as long as $e>0$.}  In other words, the mixed partial $\partial^2 \Pi(r,e) / \partial e \partial r =0$.  The implication is that, for any contract that implements a particular action, a marginal change in risk-taking away from the implemented action has no effect on the sensitivity of the payoff to effort, or vice versa. 

Continuing with the proof, set $\partial \Pi(\tilde{r},\tilde{e}) / \partial r = 0$ and re-arrange the remaining terms in \ref{eq:dLdr} to give
\begin{linenomath}
\begin{equation}
  \beta \bar{B}_r(\tilde{r}, \tilde{e}) = \beta \int \, h(\bar{u}^{(\tilde{r}, \tilde{e})}(v))  \, F_r(dv; \tilde{r}, \tilde{e}) - \mu \int \, \bar{u}^{(\tilde{r}, \tilde{e})}(v) \, F_{rr}(dv; \tilde{r}, \tilde{e}). 
\label{eq:dLdr2}
\end{equation}
\end{linenomath}
On the left of eq.~\ref{eq:dLdr2}, note that $\bar{B}_r(\tilde{r}, \tilde{e}) = dB(s(\tilde{r}, \tilde{e}))/dr  = B'(s(\tilde{r}, \tilde{e}))s_r(\tilde{r}, \tilde{e})$.  On the right, $\int \, \bar{u}^{(\tilde{r}, \tilde{e})}(v) \, F_{rr}(dv; \tilde{r}, \tilde{e}) = \frac{\partial^2}{\partial r^2 }\int \, \bar{u}^{(\tilde{r}, \tilde{e})}(v) \, F(dv; \tilde{r}, \tilde{e}) \leq 0$, because $\bar{u}^{(\tilde{r}, \tilde{e})}(v)$ maximizes the community's payoff when implementing $(\tilde{r}, \tilde{e})$.   Thus if it can be shown  that $\int \, h(\bar{u}^{(\tilde{r}, \tilde{e})}(v))  \, F_r(dv; \tilde{r}, \tilde{e}) > 0$, this together with $B'>0$, $\beta>0$, and $\mu>0$ implies that $s_r(\tilde{r}, \tilde{e})>0$, from which the result follows. \\

\noindent \textbf{Lemma:} For any action $(r,e)$, $\frac{d}{dr} \int \, h(\bar{u}^{(r,e)}(v))  \, F(dv; r, e) > 0$.

\begin{proof} The claim of the lemma is that if investigators increase their scientific risk marginally when faced with a contract that implements $(r,e)$,  then the total wage payouts will increase also.  The intuition is that because $\bar{u}^{(r,e)}(\cdot)$ implements $(r,e)$, then by RC a marginal increase in risk does not change the investigators' wage utility, that is, $\frac{d}{dr} \int \, \bar{u}^{(r,e)}(v) \, F(dv; r, e) = 0$.  However, the wages that need to be paid to create this wage-utility schedule are a convex in the wage utility, that is, $h''>0$.   Thus, because increasing risk increases the variance in the scientific value of the outcome, a marginal increase in risk increases the total wages paid.  The rest is just working out the mathematical details.

To ease the notation in the proof, suppress the $(r,e)$ superscript on $\bar{u}(v)$.  By MLRP, there is a value $v_0 > 0$ such that\footnote{Actually, the MLRP does not rule out the possibility that $f_r(v;r,e)<0$ for all $v>0$.  But if this were the case, then RC would require that $\bar{u}(v) = \bar{u}(0)$ for all $v\geq0$, which cannot implement costly effort.  So we know that $f_r(v;r,e)>0$ for some $v>0$.} 
\begin{displaymath}
f_r(v; r,e) \begin{cases} < 0 & 0 < v < v_0 \\ 
                        = 0 & v_0 \\
                        > 0 & v_0 < v. \end{cases}
\end{displaymath}
By RC, we have
\begin{linenomath}
\begin{align*}
0 & = \int_{v \geq 0} \, \left[\bar{u}(v) - \bar{u}(0) + \bar{u}(0)  \right]\, F_r(dv; r,e) \\
& = \int_{v \geq 0} \, \left[\bar{u}(v) - \bar{u}(0) \right]\, F_r(dv; r,e) + \bar{u}(0) \int_{v \geq 0} \,  F_r(dv; r,e)\\
& = \int_{v > 0} \, \left[\bar{u}(v) - \bar{u}(0) \right]\, F_r(dv; r,e) \\
& = \int_{v \in (0, v_0)} \, \left[\bar{u}(v) - \bar{u}(0) \right]\, F_r(dv; r,e) + \int_{v > v_0} \, \left[\bar{u}(v) - \bar{u}(0) \right]\, F_r(dv; r,e)   
\end{align*}
\end{linenomath}
Now, because we are only integrating over values $v>0$, write $F_r(dv; r,e) = f_r(v; r,e) \, dv$ and interpret the integrals as Riemann integrals.  Divide through by $\bar{u}(v_0) - \bar{u}(0) > 0$ to give
\begin{linenomath}
\begin{align*}
0 & = \int_{v \in (0, v_0)} \, \dfrac{\bar{u}(v) - \bar{u}(0)}{\bar{u}(v_0) - \bar{u}(0)}\, f_r(v; r,e) \, dv + \int_{v > v_0} \, \dfrac{\bar{u}(v) - \bar{u}(0)}{\bar{u}(v_0) - \bar{u}(0)} \,f_r(v; r,e) \, dv.
\end{align*}
\end{linenomath}
Now by the convexity of $h$, we have 
\begin{displaymath}
\dfrac{h(\bar{u}(v)) - h(\bar{u}(0))}{h(\bar{u}(v_0)) - h(\bar{u}(0))} < \dfrac{\bar{u}(v) - \bar{u}(0)}{\bar{u}(v_0) - \bar{u}(0)}\mbox{  for  }v < v_0
\end{displaymath}
and
\begin{displaymath}
\dfrac{h(\bar{u}(v)) - h(\bar{u}(0))}{h(\bar{u}(v_0)) - h(\bar{u}(0))} > \dfrac{\bar{u}(v) - \bar{u}(0)}{\bar{u}(v_0) - \bar{u}(0)}\mbox{  for  }v > v_0.
\end{displaymath}
Thus
\begin{displaymath}
\int_{v \in (0, v_0)} \, \dfrac{\bar{u}(v) - \bar{u}(0)}{\bar{u}(v_0) - \bar{u}(0)}\, f_r(v; r,e) \, dv  < \int_{v \in (0, v_0)} \, \dfrac{h(\bar{u}(v)) - h(\bar{u}(0))}{h(\bar{u}(v_0)) - h(\bar{u}(0))} \, f_r(v; r,e) \, dv  < 0
\end{displaymath}
and
\begin{displaymath}
0 < \int_{v > v_0} \, \dfrac{\bar{u}(v) - \bar{u}(0)}{\bar{u}(v_0) - \bar{u}(0)} \,f_r(v; r,e) \, dv < \int_{v > v_0} \, \dfrac{h(\bar{u}(v)) - h(\bar{u}(0))}{h(\bar{u}(v_0)) - h(\bar{u}(0))}\,f_r(v; r,e) \, dv 
\end{displaymath}
from which
\begin{linenomath}
\begin{align*}
0 & < \int_{v \in (0, v_0)} \, \dfrac{h(\bar{u}(v)) - h(\bar{u}(0))}{h(\bar{u}(v_0)) - h(\bar{u}(0))} \, f_r(v; r,e) \, dv + \int_{v > v_0} \, \dfrac{h(\bar{u}(v)) - h(\bar{u}(0))}{h(\bar{u}(v_0)) - h(\bar{u}(0))}\,f_r(v; r,e) \, dv \\
& = \dfrac{1}{h(\bar{u}(v_0)) - h(\bar{u}(0))} \int \, \left[ h(\bar{u}(v)) - h(\bar{u}(0)) \right] \, F_r(dv; r,e) \\ 
& = \dfrac{1}{h(\bar{u}(v_0)) - h(\bar{u}(0))}  \int \, h(\bar{u}(v)) \, F_r(dv; r,e)
\end{align*}
\end{linenomath}
and the result follows.
\end{proof}




\subsection{Numerical example}

In the numerical example, the quantities $f_e(v; r, e)/f(v; r, e)$ and $f_r(v; r, e)/f(v; r, e)$ are both linear in $v$ for $v>0$.  Specifically, we have $f_e(v; r, e)/f(v; r, e) = v/(2re^{3/2})$ while $f_r(v; r, e)/f(v; r, e) = v/(r^2\sqrt{e}) - 1/r - 1/(1-r)$ for $v>0$.  Consequently, using \ref{eq:stareqn} along with $h'(u) = 2u$ establishes that $\bar{u}^{(r,e)}(v)$ is linear in $v$ for $v>0$, and thus can be written
\begin{linenomath}
\begin{equation*}
\bar{u}^{(r,e)}(v) = \begin{cases} \bar{u}_0 & v= 0 \\ a + bv & v>0. \end{cases}
\end{equation*}
\end{linenomath}
Thus $\bar{u}^{(r,e)}(v)$ is fully determined by the triple $(\bar{u}_0, a, b)$.  The (binding) BC, EC, and RC thus create a system of three equations in three unknowns that can be solved numerically to find the contract $\bar{u}^{(r,e)}(v)$, the solution to the first sub-problem.  To solve the second sub-problem, the optimal values of $e$ and $r$ are then found by using the ``optimize'' routine in the R computing environment \cite{R}.  The ``optimize'' routine is based on the Algol 60 procedure given in \cite{brent1973algorithms}.  The second-order conditions for the second sub-problem are verified numerically at the computed optimum.

\end{document}